\documentclass[aps,pre,twocolumn,groupedaddress,showpacs]{revtex4}

\usepackage{graphicx}

\begin{document}

\title{Extinction in Lotka-Volterra model\\}

\author{Matthew Parker}
\affiliation{School of Physics and Astronomy, University of Minnesota, Minneapolis, MN 55455}

\author{Alex Kamenev\\}
\affiliation{School of Physics and Astronomy, University of Minnesota, Minneapolis, MN 55455}

\date{\today}

\begin{abstract}
Competitive birth-death processes often exhibit an oscillatory behavior.
We investigate a particular case where the oscillation cycles are marginally stable on the mean-field level.
An iconic example of such a system is the Lotka-Volterra model of predator-prey competition.
Fluctuation effects due to discreteness of the populations destroy the mean-field
stability and eventually drive the system toward  extinction of one or both species.
We show that the corresponding extinction time scales as a certain power-law of the population
sizes. This behavior should be contrasted with the extinction of models stable in the mean-field
approximation. In the latter case the extinction time scales exponentially with size.

\end{abstract}

\pacs{05.40.-a,87.23.Cc,02.50.Ey,05.10.Gg}


\maketitle\section{\label{sec:intro}Introduction}

Understanding stochastic population dynamics is an important pursuit in the biological sciences~\cite{Bartlett,Andersson,Diekmann,Daley}. Mathematical modeling of such dynamics allows for better understanding of biodiversity and species extinction. Such modeling becomes especially important because often the relevant time scales make direct measurements difficult. One of the most basic relationships that can be used to study such dynamics is the predator-prey relation. In such a system, one species reproduces by killing the other. An individual of the prey species replicates at a constant rate. Individual predators die at a constant rate and replicate only at the expense of the prey. Although the most obvious application of such a system is two organisms, the predator-prey relation can also be used to study other systems.

The original work by Lotka~\cite{Lotka} and Volterra~\cite{Volterra} showed that such a system  results in oscillations of both populations. Stochastic simulations can be used to better understand such a system. In a system without spatial degrees of freedom, the Lotka-Volterra interaction invariably results in an extinction event in which either the predator species or prey species goes extinct.
This departure from the original results can be understood as a result of stochasticity associated with the discreteness of the populations.
Such behavior has been observed in, for example, the cyclic Lotka-Volterra system~\cite{Reichenbach}. Understanding this departure from mean-field dynamics provides a challenging  problem in non-equilibrium statistical mechanics.

The unique feature of the Lotka-Volterra model is the presence of an ``accidental'' first integral of the mean-field equations of motion. As a result, all mean-field trajectories evolve on  closed orbits. These type of dynamics are marginally stable, since fluctuations in any direction are neither damped nor amplified.
Such fluctuations originate from intrinsic demographic stochasticity along with the discreteness of the populations; they lead to a slow diffusion between the mean-field orbits. Even large deviations from mean-field expectations, such as extinction, may be viewed as the accumulation of many small step fluctuations in the radial direction. This should be contrasted with reaction systems that have a stable fixed point or limiting cycle. In those system the large deviations proceed only along  very special instanton paths in the phase space~\cite{Dykman2,Elgart}.  Due to this difference, the extinction time in marginally stable systems exhibit power law dependence  on  the two populations sizes, instead of being exponentially  long as in the case of (meta)stable models.

This work gives a formulation of the problem using the Fokker-Planck equation. Using time scale separation between fast angular and slow radial motion, the inherently two-dimensional problem is reduced to a one-dimensional one. The latter is the problem of diffusion with a specific radius dependent drift. We then solve the first passage problem for this effective 1D problem and characterize the extinction probability  in the long and short time limits. We rely on extensive comparison of the analytic results with the stochastic simulations. We achieve a quantitative agreement between the two, which is in all cases is better than $5\%$ and may reach an accuracy of $0.5\%$.

Our main result may be formulated as follows: for  generic parameters and initial conditions the typical number  of cycles $C$ the system undergoes before going extinct scales as
$$C\propto N_s^{3/2}\times N_d^{-1/2}\,,$$ where $N_{d}>N_s$ are the sizes of the dominant and subdominant populations
correspondingly.  This result implies a number of surprising consequences, which were all confirmed in simulations.
For example, it predicts that a further increase of an already dominant population only {\em accelerates} the total extinction. It also shows that some very different systems behave  virtually indistinguishably vis-a-vis
extinction, if their $C$-numbers are the same. For the symmetric scenario $N_d = N_s = N$, we find
$C\propto N$ in agreement with Ref.~\cite{Reichenbach}.

The outline of this paper is as follows: in section \ref{sec:mean-field} we present the mean-field dynamics of the Lotka-Volterra system. Section \ref{sec:stochastic} presents some of the results of extensive Monte Carlo simulations. An analytic approach to understanding the problem is presented in section \ref{sec:analytics}. Finally, the results are discussed in section \ref{sec:discussion}.

\section{\label{sec:mean-field}Mean-Field Theory}

In a basic predator-prey system, there are two populations. The predator species has a death rate, $\sigma$, and the prey has a birth rate, $\mu$. In addition, there is a cross reaction where a predator consumes a member of the prey population in order to reproduce.  This occurs at rate $\lambda$. The reaction scheme  can be summarized as follows:
\begin{equation}\label{eqn:reactions}
F\stackrel{\sigma}{\rightarrow} 0\,; \qquad R\stackrel{\mu}{\rightarrow} 2R\,; \qquad F+R\stackrel{\lambda}{\rightarrow} 2F\,,
\end{equation}
where $F$ signifies a predator (``fox'') and $R$ signifies a prey individual (``rabbit'').

In the mean-field approximation one neglects the discreteness of the populations and models the system with deterministic rate equations. If $q_1$ and $q_2$ are taken to be continuous variables representing the predator and prey populations, the dynamics of these two variables are given by the following  equations
\begin{eqnarray}\label{eqn:MFrates}
\dot{q_1}&=&-\sigma q_1+\lambda q_1 q_2\, ,\nonumber\\
\dot{q_2}&=&\mu q_2-\lambda q_1 q_2\, .
\end{eqnarray}
The rate of change of the predator population contains a death term proportional to the predator population and a birth term proportional to the size of both populations.  Likewise, the rate of change of the prey population has a birth rate proportional to the prey population and a death term proportional to both. Some features of these dynamics are immediately evident. There are three fixed points. These correspond to $(q_1 , q_2) = (0,0),(0,\infty)$, and  $(\mu/\lambda$,$\sigma/\lambda)$.  The first point corresponds to the trivial case of extinction of both species. The second fixed point is the result of predator extinction and the prey population growing exponentially. The third is the coexistence fixed point, where the stable populations of the predator and prey are $N_1=\mu/\lambda$ and $N_2=\sigma/\lambda$.

\begin{figure}
\includegraphics[width=3in]{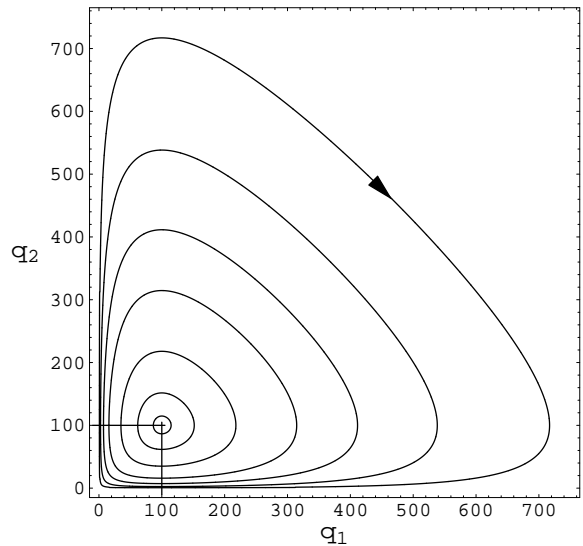}
\caption{\label{fig:Equig_Old} Orbits of constant $G=(0.01,0.1,0.4,1,1.7,2.7,4.2)$ in units of $\sqrt{\sigma \mu}$. The evolution proceeds clockwise around the mean-field fixed point of $N_1=N_2=100$.}
\end{figure}

For a given  initial condition, the populations evolve along a closed orbit in predator-prey space. The orbits are closed due to the existence of an ''accidental'' integral of motion in the mean-field equations of motion  (\ref{eqn:MFrates})
\begin{equation}
\label{eqn:Ginitial}
G = \lambda q_1 - \mu - \mu\ln{(q_1\lambda/\mu)} +\lambda q_2 - \sigma - \sigma\ln{(q_2\lambda/\sigma)}.
\end{equation}
The definition of $G$ is chosen such that $G=0$ corresponds to the coexistence fixed point, while $G\to \infty$ corresponds to large amplitude cycles closely approaching the two axes. Figure \ref{fig:Equig_Old} shows orbits for various values of the integral of motion, $G$, for the case  $N_1=N_2=100$. The presence of the integral of motion makes all cycles
marginally stable. Indeed, a small fluctuation may shift the system from one orbit to a neighboring one. Since
the new orbit is also a stable solution of the mean-field equations of motion, there is neither a restoring force, trying to compensate for the fluctuation, nor amplification of the fluctuation.

\begin{figure}
\includegraphics[width=3in]{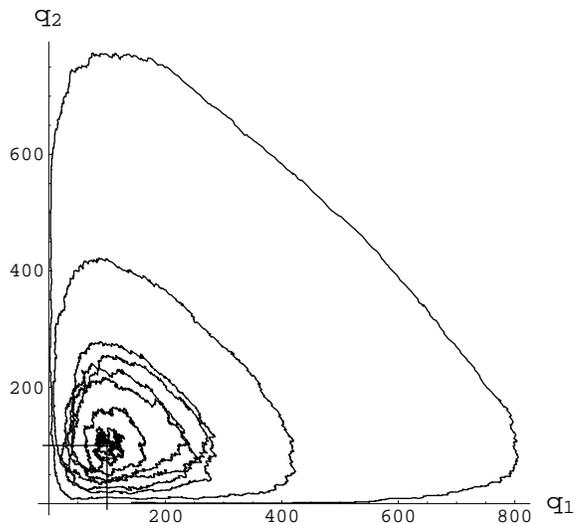}
\caption{\label{fig:Stoc} Typical run of the stochastic simulation of the model (\ref{eqn:reactions}) for $N=100$ and $\epsilon=1$.}
\end{figure}

At small $G$, the mean-field orbits are ellipses. The frequency of these elliptical orbits approaches a constant value, $1/\sqrt{\sigma \mu}$, as $G$ approaches zero. This provides a natural time scale for the problem. By rescaling time to be measured in these units, it is possible to reduce the number of parameters in the problem from the original three reaction rates to two parameters, which are convenient to choose as
\begin{equation}\label{eqn:N_eps}
N = \frac{\sqrt{\mu\sigma}}{\lambda}=\sqrt{N_1N_2}\,; \qquad \epsilon = \sqrt{\sigma\over \mu}=\sqrt{N_2\over N_1}\,.
\end{equation}
Here $N$ represents an effective system size, while $\epsilon$ represents the asymmetry between the predator and prey populations. Throughout this paper we shall be interested in the limit of large system size $N\gg 1$. By the reasons
explained below the asymmetry parameter $\epsilon$ will be restricted to the interval $N^{-1/2}< \epsilon < N^{1/2}$.

\section{\label{sec:stochastic}Stochastic Simulations}

The mean-field approximation fails to accurately portray the actual evolution of the reaction system, Eq.~(\ref{eqn:reactions}). As previously mentioned, the mean-field solution does not take into account the stochasticity associated with individual birth-death events or the discreteness of the populations. Results from Monte Carlo simulations of this reaction scheme demonstrate the failure of the mean-field.

Stochastic simulations were done as follows. The initial populations were taken to be at the coexistence fixed point. Time was discretized into small steps of size $\delta t$. The time step, $\delta t$, was chosen so that the probability of having any change in population size during $\delta t$ was small, i.e.  $\delta t \ll 1/N\sqrt{\sigma \mu}$. For each time step, the number of prey births and predator deaths was calculated randomly from a binomial distribution with success rates $\mu \delta t$ and $\sigma \delta t$ respectively. The number of consumption reaction events was calculated from a binomial distribution based on $\lambda \delta t$ and the number of predator/prey pairs. This was repeated until one of the populations went extinct. Figure \ref{fig:Stoc} shows an example of such a simulation. As in the mean-field case, the system rotates clockwise about the coexistence fixed point. As time progresses, however, the system unwinds from the fixed point, eventually hitting either the $q_1$ or $q_2$ axis. From there, the system rapidly progresses toward one of the extinction fixed points. For a typical simulation, the system rotates around the fixed point many times before going extinct.

\begin{figure}
\includegraphics[width=3in]{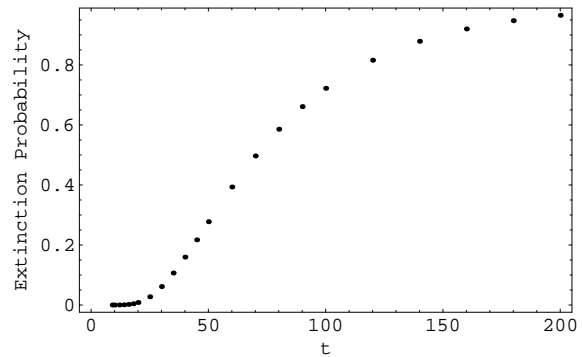}
\caption{\label{fig:tauhist} Extinction probability in time $t$ from  $10^5$ simulation trials ($N=100$, $\epsilon = 1$). Time is in units of $1/\sqrt{\sigma \mu}$. }
\end{figure}

\begin{figure}
\includegraphics[width=3in]{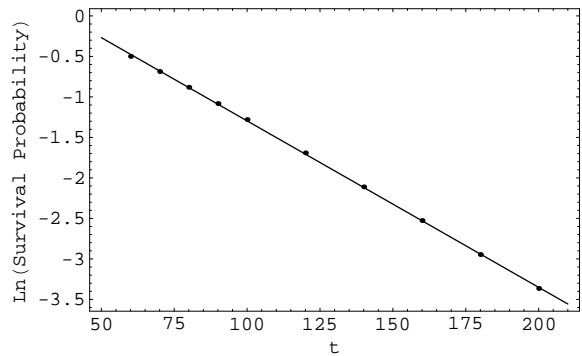}%
\caption{\label{fig:hightcalc} Logarithm of the survival probability at long times for the data presented in Fig.~\ref{fig:tauhist}.}
\end{figure}

\begin{figure}
\includegraphics[width=3in]{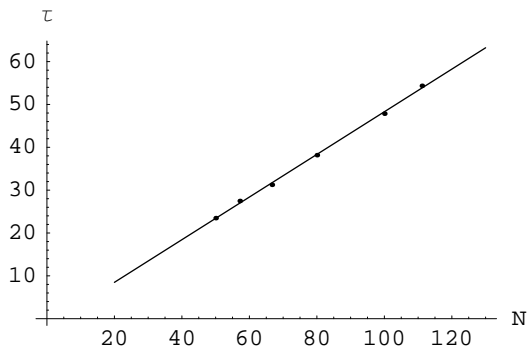}%
\caption{\label{fig:tauvN} Time constant $\tau_l$ of exponential decay, Eq.~(\ref{eqn:surv-prob}), versus $N$ for $\epsilon=1$. }
\end{figure}

Since such a simulation invariably ends in the extinction of one or both species, it is interesting to analyze the chance of the system being dead as a function of time. For a given set of initial conditions, it is possible to determine this extinction probability by repeatedly running a stochastic trial. Figure \ref{fig:tauhist} shows the result of 100,000 stochastic simulations using the conditions of the simulation presented in Fig.~\ref{fig:Stoc}. As could be expected, at short time scales there is very little chance for the system to be extinct. As $t$ grows the probability of extinction approaches unity.
At long time scales, the convenient quantity for calculation is not the extinction probability, but the survival probability, this being the likelihood of the system still being alive at a given time, $t$. The logarithm of the survival probability appears to be linear in time, Fig.~\ref{fig:hightcalc}. As a result, the survival probability is decaying exponentially with a characteristic time $\tau_{l}$,
\begin{equation}
\label{eqn:surv-prob}
P_{surv}(t) = 1 - P_{ext}(t) \propto e^{-t/\tau_{l}}\,.
\end{equation}
Figure \ref{fig:tauvN} shows the dependence of $\tau_{l}$ on $N$. One observes a linear growth of the characteristic time
$\tau_l$  with increasing $N$ at $N\gg 1$. This agrees with the results observed by Reichenbach, et al.~\cite{Reichenbach} for the cyclic Lotka-Volterra system. This linear dependence suggests the following representation for $\tau_{l}(N,\epsilon)$ in the limit $N\gg 1$:
\begin{equation}\label{eqn:tlong}
\tau_{l}(N,\epsilon)=\frac{N}{E_0(\epsilon)}\, ,
\end{equation}
where the rescaled extinction rate, $E_0$, depends only on the asymmetry, $\epsilon$, but not on $N$. The fit of Fig.~\ref{fig:tauvN} gives an observed value of $E_0(1) = 2.05$.

\begin{figure}
\includegraphics[width=3in]{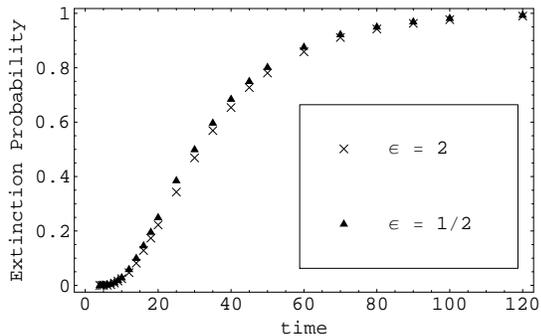}
\caption{\label{fig:EP2} Extinction probabilities for $\epsilon=2$ and $\epsilon=1/2$; N=100.}
\end{figure}

\begin{figure}
\includegraphics[width=3.5in]{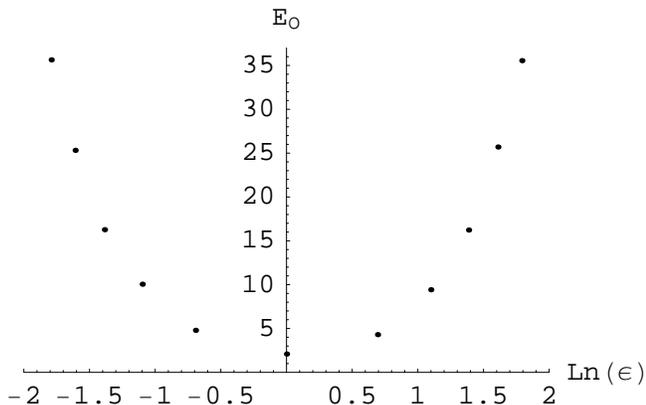}
\caption{\label{fig:E0_v_epsilon} Plot of $E_0$ vs. $\ln \epsilon$ ; $N=100$.}
\end{figure}

We focus now on the role of  the asymmetry parameter $\epsilon$. Figure \ref{fig:EP2} plots the extinction probabilities versus time for $\epsilon=2$ and $\epsilon=1/2$. The plots show virtually identical behavior. In particular, the similarity in the long time decay suggests a symmetry in $E_0$ between $\epsilon$ and $1/\epsilon$. Figure \ref{fig:E0_v_epsilon} shows a plot of the observed $E_0$ vs. the logarithm of $\epsilon$ in stochastic simulation, confirming  that
\begin{equation}
E_0(\epsilon)=E_0(1/\epsilon).
\end{equation}
The minimum of $E_0$ corresponds to $\epsilon=1$. Away from this point one observes
$E_0(\epsilon)=0.97\big(\mbox{max}\{\epsilon,1/\epsilon\}\big)^2$.

\begin{figure}
\includegraphics[width=3in]{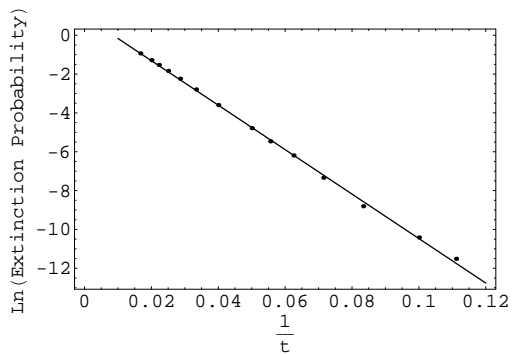}%
\caption{\label{fig:lowtcalc} Logarithm of the extinction probability at short times for the data of Fig.~\ref{fig:tauhist}.}
\end{figure}

Unlike the long times, the short time behavior is not universal and depends on the choice of the initial conditions.
For the initial populations chosen close to the coexistence fixed point $(N_1,N_2)$
it is exceedingly unlikely for the system to drift all the way  to extinction in a short time interval. Plotting the logarithm of the extinction probability vs. {\em inverse} time shows a linear dependence at small times. This can be seen for $\epsilon=1$ in Fig.~\ref{fig:lowtcalc}. Extinction probability in small times is thus exponentially small and appears to have a functional form of
\begin{equation}\label{eqn:pextprop}
P_{ext} \propto e^{-\tau_{s}/t}.
\end{equation}
As in the long time limit, the time constant $\tau_s(N,\epsilon)$ scales linearly with  $N$ for $
N\gg 1$. It is thus convenient to parameterize  $\tau_{s}(N,\epsilon)$ by an $N$-independent quantity $X_0(\epsilon)$ as
\begin{equation}
\tau_{s}(N,\epsilon) = N\frac{X_0^2(\epsilon)}{4}
\end{equation}
This parameterizations  of $\tau_s$ puts Eq.(\ref{eqn:pextprop}) in a form that is reminiscent of a standard diffusion propagator. As was the case for $E_0(\epsilon)$, $X_0(\epsilon)$ also shows symmetry between $\epsilon$ and $1/\epsilon$, i.e.
$X_0(\epsilon) = X_0(1/\epsilon)$.
From the  simulations   one observes  $X_0(1)=2.09$ at the maximum.

\section{\label{sec:analytics}Analytic Approach}
\subsection{\label{ssec:mastereq}Master and Fokker-Planck Equations}

The full behavior of the reaction model (\ref{eqn:reactions}) can be analyzed by employing a probability distribution and studying its dynamics. Define such a probability distribution $P(m,n;t)$ as the probability of the system having $m$ predators and $n$ prey at time $t$, where $m$ and $n$ are both non-negative integers. This yields the following master equation for the reaction scheme of Eq.~(\ref{eqn:reactions})
\begin{eqnarray}
\label{eqn:Master}
&&\partial_t{P}(m,n;t)=\sigma [(m +1)P(m+1,n)-m P(m,n)]\nonumber\\
&&+\, \mu [(n -1)P(m,n -1)\!-\! n P(m,n)]\\
&&+\,\lambda [(m -1)(n+1)P(m -1,n +1)-mn P(m,n)].\nonumber
\end{eqnarray}
This equation can be rewritten using integer shift operators, defined as
\begin{equation}
\label{eqn:shift}
\hat{E}_{k,l} = e^{k \partial_m+l \partial_n}\,,
\end{equation}
to give
\begin{eqnarray}
\partial_t{P}(m,n;t)&=&\sigma (\hat{E}_{1,0}-1)m P(m,n;t)\nonumber\\
&+&\mu (\hat{E}_{0,-1}-1)n P(m,n;t)  \\
&+&\lambda (\hat{E}_{-1,1}-1)mn P(m,n;t)\,.\nonumber
\end{eqnarray}

\begin{figure}
\includegraphics[width=3in]{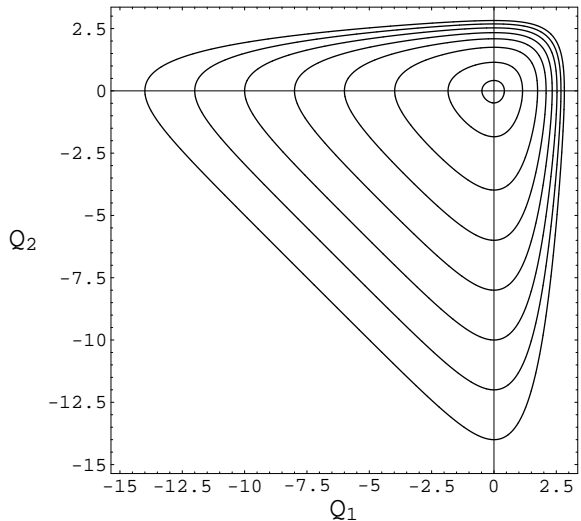}
\caption{\label{fig:equig_newc} Orbits of constant $G$ in the new coordinates}
\end{figure}

An important distinction of the models with marginally stable cycles is that a large deviation (such as an extinction) may proceed in a sequence of small steps. A  small fluctuation leads to a mean-field like evolution along a new stable orbit until another  small fluctuation shifts the orbits again, {\em etc}. As a result, a path to extinction is a random diffusion in population space.
This should be contrasted with models with stable limiting cycles or an attracting fixed point~\cite{Kamenev,Dykman}, where  small fluctuations  do not accumulate and extinction proceeds only along a very specific (instanton) trajectory. On the technical level this observation implies that the gradients $\partial_{m,n}$ may be considered as small $\sim 1/N$ (this is usually not the case on the instanton trajectory~\cite{Gaveau,Doering,Dykman2,Elgart,Assaf,Assaf2}) and thus the shift operators (\ref{eqn:shift}) may be expanded up to the second order. This procedure leads to the Fokker-Planck (FP) equation, which in the present context is justified by the Van-Kampen expansion over the system size $N$~\cite{Kampen}.
Proceeding this way, one finds
\begin{eqnarray}\label{eqn:FP}
\partial_t{P}=\sigma \left[\partial_{q_1}+\frac{1}{2}\partial^{2}_{q_1}\right]q_1 P+\mu \left[-\partial_{q_2}+\frac{1}{2}\partial^{2}_{q_2}\right]q_2 P\nonumber\\
+\lambda \left[\partial_{q_2}+\frac{1}{2}\partial^{2}_{q_2}-\partial_{q_1}+
\frac{1}{2}\partial^{2}_{q_1}-\partial_{q_1} \partial_{q_2}\right]q_1 q_2 P.
\end{eqnarray}
This equation along with Eq.(\ref{eqn:MFrates}) suggest a change of variable such that the new variables $Q_i\sim \ln q_i$.  This is accomplished through the following transformation:
\begin{equation}\label{eqn:coordtrans}
q_1 = \frac{\mu}{\lambda}\,e^{\sqrt{\frac{\sigma}{\mu}}\,{Q_1}} \qquad q_2 = \frac{\sigma}{\lambda}\,e^{\sqrt{\frac{\mu}{\sigma}}\,{Q_2}}
\end{equation}
These variables present some advantage over the initial ones. Extinction events now occur at $Q_1=-\infty$ or $Q_2=-\infty$ instead of at $q_1=0$ or $q_2=0$. The coexistence fixed point has been moved to the origin. As part of this transformation, time is rescaled into the problem's natural units, $1/\sqrt{\sigma \mu}$. The FP equation no longer depends on the three reaction rates; it depends only on $N$ and $\epsilon$.
In the new coordinates the mean-field integral of motion takes  the form
\begin{equation}
G=\frac{1}{\epsilon}(e^{\epsilon Q_1 }-1)-Q_1 +\epsilon(e^{Q_2/ \epsilon}-1)-Q_2.
\end{equation}
It provides a natural radial coordinate. The coexistence fixed point is $G = 0$, while extinction corresponds to $G=\infty$. Figure \ref{fig:equig_newc} shows mean-field orbits in the transformed coordinate system. Larger orbits correspond to larger values of $G$.
The most essential advantage of the new variables is that the mean-field equations (\ref{eqn:MFrates}) acquire the
Hamiltonian structure, where $Q_1$ and $Q_2$ form a canonical  pair
\begin{eqnarray}\label{eqn:MFdyn}
\dot{Q_1}&=&-1+e^{Q_2/\epsilon} = \partial_{Q_{2}} G\, ;\nonumber\\
\dot{Q_2}&=&1-e^{\epsilon Q_1}=- \partial_{Q_1} G\,.
\end{eqnarray}
Since $G(Q_1,Q_2)$ serves as the Hamiltonian, it is manifestly conserved on the solutions of the mean-field equations of motions.

The probability distribution is transformed in the new coordinate system so as to include the Jacobian of the transformation
\begin{equation}
W(Q_1,Q_2;t) = q_1 q_2 P(q_1,q_2;t),
\end{equation}
where $q_1$ and $q_2$ are substituted from Eq.~(\ref{eqn:coordtrans}).
In the new coordinates, the Fokker-Plank equation~(\ref{eqn:FP}) becomes
\begin{equation}\label{eqn:continuity}
\partial_t{W}=-\vec{\nabla}\cdot \vec{J},
\end{equation}
where the divergence is defined as
\begin{equation}
\vec{\nabla}=(\partial_{Q_1},\partial_{Q_2}).
\end{equation}
The probability current in Eq.~(\ref{eqn:continuity}) consists of two parts
\begin{equation}
\vec{J} = \vec{J}^{MF}+\vec{J}^{D}\,.
\end{equation}
The mean-field motion along the orbits of constant $G$ is due to $\vec{J}^{MF}$, see Eq.~(\ref{eqn:MFdyn}), while the  radial diffusion between the orbits is due to $\vec{J}^{D}$. The mean-field current is given by
\begin{equation}
J_1^{MF} = (-1+e^{Q_2/\epsilon})W = (\partial_{Q_2} G) W\,;
\end{equation}
\begin{equation}
J_2^{MF} = (1-e^{\epsilon Q_1})W = -(\partial_{Q_1} G) W\,.
\end{equation}
The diffusive  current is found from Eq.~(\ref{eqn:FP}) as
\begin{equation}
\label{eqn:JD1}
J_1^D=-\frac{1}{2N}\left[ (e^{-\epsilon Q_1}+e^{\frac{Q_2}{\epsilon}-\epsilon Q_1})\partial_{Q_1} W-\partial_{Q_2} W\right]\,;
\end{equation}
\begin{equation}
\label{eqn:JD2}
J_2^D=-\frac{1}{2N}\left[ (e^{-\frac{Q_2}{\epsilon}}+e^{\epsilon Q_1-\frac{Q_2}{\epsilon}})\partial_{Q_2} W-\partial_{Q_1} W\right].
\end{equation}
The diffusive current is suppressed by a factor of $N$ relative to the mean-field current. Provided the system is sufficiently large, i.e. $N \gg 1$, this means that the angular motion should be much faster than the radial motion.

\subsection{\label{ssec:oned}Reduction to One Dimension}

As mentioned, the mean-field constant, $G$, provides a natural radial coordinate. The corresponding angular coordinate evolves far faster than the radial one. This time-scale separation represents an opportunity to turn this two-dimensional problem into a one-dimensional one. The method used has been successfully employed in the analysis of spin-torque switching~\cite{Apalkov}. Since $Q_1$ and $Q_2$ form a canonical pair on the mean-field level, it is possible to transform them into action-angle variables $(I,\alpha)$ where the action is an integral of the mean-field motion, i.e. $G=G(I)$. It is more convenient to use $G$ itself, rather than the canonical action, $I$, as the radial variable. One should be aware though that the change of variables $(Q_1,Q_2)\to (G,\alpha)$ involves a Jacobian, discussed below.

We shall assume now that due to the fast angular motion the probability distribution $W(G,\alpha;t)$ rapidly equilibrates in the angular direction and at the long time scales depends on the radial variable only, i.e.
$W=W(G;t)$.  Under this assumption it is possible to eliminate the angular dependence from Eq.~(\ref{eqn:continuity}) by integrating the continuity equation (\ref{eqn:continuity}) over the area of a mean-field orbit with a fixed $G$
\begin{equation}
 \int\!\!\int_G \partial_t{W}(G) dQ_1 dQ_2= - \int\!\!\int_G \vec{\nabla}\cdot\vec{J}\, dQ_1 dQ_2
\end{equation}
We apply the divergence theorem to the right hand side and change the coordinates of integration on the left hand side. The unit vector, $\hat{n}$, is normal to the line of constant $G$, and $dl$ is an infintismal length along the $G$ orbit
\begin{equation}
\int\!\!\! \int_G \! \partial_t{W}(G) \left |\frac{\partial Q_1}{\partial G}\frac{\partial Q_2}{\partial \alpha}-\frac{\partial Q_2}{\partial G}\frac{\partial Q_1}{\partial \alpha}\right| dG d\alpha =\! - \!\oint_G \vec{J} \cdot \hat{n} dl
\end{equation}
The Jacobian can be reduced to $|dt/d\alpha|$ due to the Hamiltonian equations on $Q_1$ and $Q_2$, see Eq.(\ref{eqn:MFdyn}), leading to
\begin{equation}
\int \int_G \partial_t{W}(G)\left| \frac{dt}{d\alpha}\right|\, d\alpha dG = -\oint_G \vec{J} \cdot \hat{n} dl.
\end{equation}
Integration over $\alpha$ gives the period of the mean-field revolution around the orbit $T(G)$
\begin{equation}
\int_0^G\! T(G)\, \partial_t{W}(G)dG = -\oint_G \vec{J} \cdot \hat{n} \, dl\,.
\end{equation}
Finally, differentiation with respect to $G$ yields the radial  FP equation
\begin{equation}
T(G)\,  \partial_t{W}(G) = \frac{\partial}{\partial G}\left[-\oint_G \vec{J} \cdot \hat{n} dl \right]\,.
\end{equation}

We now wish to understand the integral of the current in this equation. The mean-field portion of the current, $\vec{J}^{MF}$, is perpendicular to $\hat{n}$. It therefore makes no contribution to this integral
\begin{equation}
\vec{J}^{MF} \cdot \hat{n} = 0\,.
\end{equation}
This leaves integration over the diffusive current $\vec{J}^D$, which is first-order in derivatives and proportional to $1/N$. Since we are assuming that $W$ is independent of $\alpha$, this implies that the diffusive current should be proportional to $\partial_G W$. The integral along the orbit may then be written in the form
\begin{equation}
\label{eqn:DGdef}
-\oint_G \vec{J}^D\cdot\hat{n}\, dl = \frac{1}{N}\, D(G)\, \frac{\partial W}{\partial G}\,,
\end{equation}
which is in essence the definition of the effective diffusion parameter, $D(G)$. The resulting FP equation takes the form
\begin{equation}\label{eqn:1DFP}
T(G)\, \partial_t{W}(G)=\frac{\partial}{\partial G}\left[\frac{1}{N}\,D(G)\,\frac{\partial W(G)}{\partial G}\right]\,,
\end{equation}
where the two functions $D(G)$ and $T(G)$ may be evaluated for any mean-field orbit $G$.  They both are independent of $N$,  but do depend on $\epsilon$. We evaluate  both these quantities  in the Appendix. At small $G\ll 1$,  the period is $T=2\pi$ (which was our initial motivation of choosing these units of time) and  $D(G)\propto G$. Introducing variable $R=\sqrt{G}$, so $\partial_G=(1/2R)\partial_R$, one reduces the right hand side of  Eq.~(\ref{eqn:1DFP}) to be
$\sim R^{-1}\partial_R(R\partial_R W)$, which is the radial part of the standard 2d diffusion equation. At large $G$ the period grows linearly, $T(G)\sim G$, while the diffusivity $D(G)$ grows exponentially. The latter fact is due to the two sharp maxima in the
current $\vec J^D$ which take place around the two ``arms'' of the mean-field orbits, Fig.~\ref{fig:equig_newc}, along the negative directions of the two axis. Both $T(G)$ and $D(G)$ are symmetric with respect to $\epsilon\to 1/\epsilon$, rendering the corresponding symmetry to all the results  obtained upon averaging over the angular motion.

The boundary conditions  for Eq.~(\ref{eqn:1DFP}) are as follows: since the system is incapable of moving from the extinct state to a live one, there is an absorbing boundary condition as $G\rightarrow\infty$. Since $G$ is a radial coordinate, the current must disappear at $G=0$. This gives the following boundary conditions
\begin{eqnarray}
\lim_{G\rightarrow \infty}W(G) &=& 0\nonumber \,;\\
G\, \frac{\partial W(G)}{\partial G}\Big|_{G=0} &=& 0\,,
                                                 \label{eqn:boundary-cond}
\end{eqnarray}
where we have employed $D(G)\sim G$ at $G\to 0$.
At large $G$, the diffusivity $D(G)$ grows exponentially, see Appendix. This allows the system to diffuse to $G=\infty$ in finite time which suggests that the spectrum of Eq.~(\ref{eqn:1DFP}) with the boundary conditions (\ref{eqn:boundary-cond}) is discrete, despite the equation being formulated on the infinite interval $G\in[0,\infty)$. To see this fact most clearly it is useful to introduce one more change of variables.

\subsection{\label{ssec:semiclassics}Reduction to the finite interval}

\begin{figure}
\includegraphics[width=3in]{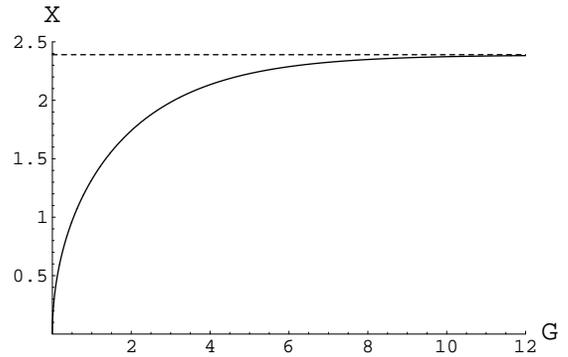}
\caption{\label{fig:X_G} $X(G)$ for $\epsilon = 1$. At $G\to\infty$ the function converges to $X_0 = 2.39$}
\end{figure}

To motivate the new variable, let us perform  a semi-classical analysis of Eq.~(\ref{eqn:1DFP}). To this end we  represent the probability distribution as
\begin{equation}\label{eqn:WexpS}
W(G;t) \propto e^{-N S(G;t)}
\end{equation}
and assume for the moment that $S\sim O(1)$ (this assumption is indeed true at short time scales, but breaks down
at $t\sim N$). Then  to the leading order in $N$, Eq.~(\ref{eqn:1DFP}) becomes
\begin{equation}
\label{HJ}
-T(G)\, \partial_t{S}(G;t) = D(G)\left( \frac{\partial S(G;t)}{\partial G} \right)^2\, ,
\end{equation}
which may be viewed as the Hamilton-Jacobi equation with the Hamiltonian
\begin{equation}
H(G,P_G) = \frac{D(G)}{T(G)}P_G^2.
\end{equation}
It is convenient to make a canonical transformation from $(G,P_G) \rightarrow (X,P_X)$ such that the Hamiltonian in the new coordinates takes the  form $H(X,P_X)=P_X^2$. This is accomplished by
\begin{eqnarray}\label{eqn:Xdef}
X &=& \int_0^G \sqrt{\frac{T(G')}{D(G')}}\, \, dG'\,;\\
\nonumber
P_X &=& \sqrt{\frac{D(G)}{T(G)}}\, P_G\, .
\end{eqnarray}
Unlike $G$, the new radial variable $X$ is bounded. Indeed, due to the exponential growth of $D(G)$ at large $G$ the
integral in Eq.~(\ref{eqn:Xdef}) converges.  For the case $\epsilon=1$, $X(G)$ is plotted in Fig.~\ref{fig:X_G}, exhibiting convergence to $X_0=2.39$.

\begin{figure}
\includegraphics[width=3in]{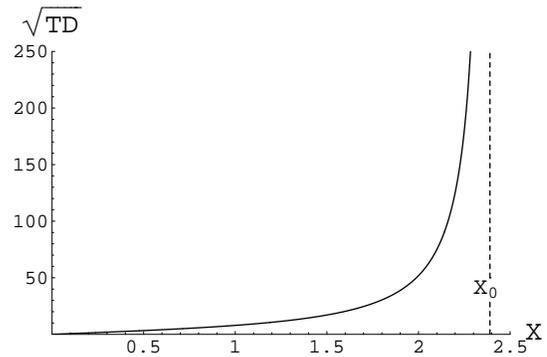}
\caption{\label{fig:td_x} Numerically calculated $\sqrt{T(X)D(X)}$ for $\epsilon = 1$. The function diverges at $X = X_0=2.39$}
\end{figure}

Although we motivated the change of variables $G\to X(G)$ by the semiclassical analysis of
Eq.~(\ref{eqn:1DFP}), one may go back to the full FP equation~(\ref{eqn:1DFP}) and perform the
variable change (\ref{eqn:Xdef}) exactly. The result is
\begin{equation}\label{eqn:wx}
\partial_t{W} = \frac{1}{N}\, \frac{1}{\sqrt{D(X)T(X)}}\,\frac{\partial}{\partial X}\left(\sqrt{D(X)T(X)}\,  \frac{\partial{W}}{\partial X}\right)\,.
\end{equation}
This equation is defined on the finite interval
$X\in [0,X_0]$.  The boundary conditions (\ref{eqn:boundary-cond}) take the form
\begin{equation}
\label{eqn:boundary-cond-new}
W(X_0;t)=0\,;\qquad X\, \frac{\partial W(X;t)}{\partial X}\Big|_{X=0}=0\,,
\end{equation}
where we take into account that $\sqrt{D(X)T(X)}\sim X$ at $X\to 0$. Equation (\ref{eqn:wx}) has framed the problem so that it no longer depends on $T$ and $D$ separately, but rather only on $\sqrt{TD}$ as well as the constant $X_0$. A plot of $\sqrt{T(X)D(X)}$ is shown in Fig.~\ref{fig:td_x}.


\subsection{Long time dependence of the extinction probability}

The long time behavior of the system can be analyzed via an eigenvalue problem on Eq.~(\ref{eqn:wx}) with the boundary conditions (\ref{eqn:boundary-cond-new}).
Since $X_0$ is finite, its spectrum is discrete. We shall look thus for a solution of the FP equation in the form
\begin{equation}
W(X;t)=\sum_n a_n w_n (X)\, e^{-E_n t/N}
\end{equation}
where $a_n$ are constants depending on initial conditions and $w_n(X)$ and $E_n$ are solutions of the following eigenvalue problem
\begin{equation}
\hat{H} w_n(X) = E_n w_n(X)\, .
\end{equation}
Here the operator $\hat H$ is defined as
\begin{equation}
\label{eqn:Hoperator}
\hat{H} = \frac{-1}{\sqrt{D(X)T(X)}}\,\frac{\partial}{\partial X}\left(\sqrt{D(X)T(X)}\, \frac{\partial}{\partial X}\right).
\end{equation}
The $N$ dependence has been explicitly removed from the eigenvalues. The only remaining dependence of $E_n$ is on the asymmetry $\epsilon$. The survival probability is given by $P_{surv}(t)=\int^{X_0}_0dX\,W(X;t)$. At long time scales, the only contributing eigenstate is the one with the smallest eigenvalue, $E_0$, and thus the survival probability   decays with the characteristic time scale $\tau_l=N/E_0(\epsilon)$,
\begin{equation}
P_{surv}(t) \propto e^{-E_0(\epsilon) t/N}\,.
\end{equation}
This is what was observed in Fig.~\ref{fig:lowtcalc}. From the fit of this figure, the observed value from stochastic simulation is $E_0(1)=2.05$.

\begin{figure}
\includegraphics[width=3in]{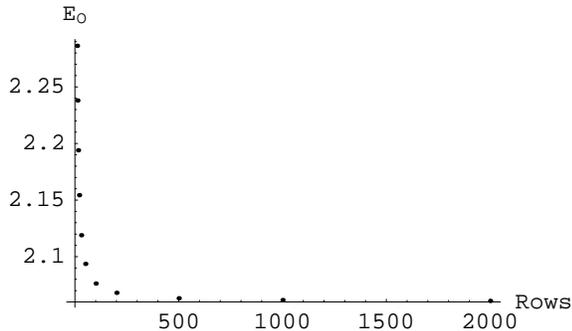}
\caption{\label{fig:E0_v_Rows} $E_0$ numerically calculated via matrix diagonalization vs. number of rows in matrix}
\end{figure}

In order to compare it with our analytical approach  we discretize the interval $[0,X_0]$ and represent the linear operator $\hat H$, Eq.~(\ref{eqn:Hoperator}), as a matrix. We then diagonalize it numerically, using expressions  for $T$ and $D$ from the Appendix.
The lowest eigenvalue, $E_0$,  converges quite rapidly with decreasing discretization step of $X$; 100 rows is sufficient to calculate $E_0$ to within $1\%$ of the convergent value. Figure \ref{fig:E0_v_Rows} shows the convergence of $E_0$ as the number of matrix rows is increased. This procedure gives for the operator $\hat H$ a lowest eigenvalue of $E_0(1)=1.95$, which agree with the Monte Carlo simulations within $5\%$ accuracy.
The discrepancy  can be  reduced even father by taking into account the finite size effect. For finite values of $N$, it is not necessary  to diffuse all the way to $G=\infty$, but only to a value of $G$ which corresponds to a single remaining  individual at a minimum of one of the populations. At this point the fluctuations will drive the system to extinction with probability close to one. Such a cutoff  $G_{ext}$ may be estimated, using Eq.~(\ref{eqn:Ginitial}), as
\begin{equation}
\label{eqn:Gextinction}
G_{ext} =\left\{ \begin{array}{ll}
\epsilon^{-1}(\ln (N/\epsilon)-1)\,; &\,\,\,\,\, \epsilon>1\,,\\
\epsilon(\ln (N \epsilon)-1)\,; &\,\,\,\,\, \epsilon<1\,.
\end{array}\right.
\end{equation}
For our simulations with $N=100$ and $\epsilon=1$ this gives $G_{ext}=3.62$.  Integrating Eq.~(\ref{eqn:Xdef}) only up to $G_{ext}$ gives $X_{0}=2.08$ (instead of $X_0=2.39$ for the infinite interval). Using this truncated $X_0$ as the cut-off for the matrix diagonalization  of $\hat H$ gives $E_0(1)=2.06$, within $0.5\%$ of stochastic simulations. Since $G_{ext}$ depends on the system size $N$ only logarithmically, it is computationally unfeasible to eliminate the finite size effect in stochastic modeling.

\begin{figure}
\includegraphics[width=3in]{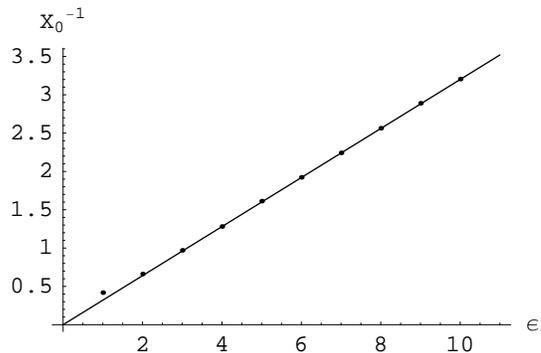}
\caption{\label{fig:X0epsilonfit} Function $X_0^{-1}(\epsilon)$.}
\end{figure}

\begin{figure}
\includegraphics[width=3in]{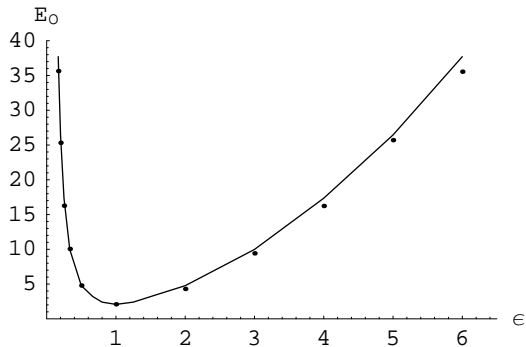}
\caption{\label{fig:E0epsilonfit} Function $E_0(\epsilon)$; dots - results of stochastic simulation, full line - operator $\hat H$ diagonalization.}
\end{figure}

We turn now to the $\epsilon$ dependence away from $\epsilon=1$. Let us discuss the $\epsilon>1$ case, i.e. $N_2>N_1$
prey dominated system (the  predator dominated scenario may be analyzed in the same way). In this case it is almost certain that predators go extinct first. This is because the diffusive current toward predator extinction, Eq.~(\ref{eqn:iDG}), is exponentially bigger than that toward prey extinction. Neglecting the latter, one observes
from  Eq.~(\ref{eqn:iDG}) that $D=D(\epsilon G)\sim e^{\epsilon G}$. This implies that the integration interval contributing to $X_0$, Eq.~(\ref{eqn:Xdef}), is effectively limited to $0<G\lesssim 1/\epsilon<1$. In this interval the period $T(G)\approx \mbox{const}$. Rescaling variables in  Eq.~(\ref{eqn:Xdef}) as $\epsilon G\to G$, one
finds that $X_0(\epsilon)\sim 1/\epsilon$ for $\epsilon >1$, Fig.~\ref{fig:X0epsilonfit}. Correspondingly, $D(X)=D(\epsilon X)$ and after rescaling $\epsilon X\to X$ in Eq.~(\ref{eqn:Hoperator}) one observes that $
E_n(\epsilon)\sim \epsilon^2$. Finally, one finds for the characteristic extinction time of an asymmetric model
\begin{equation}\label{eqn:taul-asymetric}
    \tau_l=\frac{N}{E_0(\epsilon)}=1.03\,  N \big(\mbox{max}\{\epsilon, 1/\epsilon\} \big)^{-2}\,.
\end{equation}
where the numerical factor is obtained through numerical diagonalization of the
$\hat H$ operator. Figure \ref{fig:E0epsilonfit} plots the observed values from Monte-Carlo simulation fit with our analytic prediction. Since the approach relies on the separation of time scales between the fast angular motion and the slow radial one, it requires $\tau_l> 1$. This leads to the restriction on the asymmetry parameter:
$N^{-1/2} <\epsilon<N^{1/2}$, stated in section \ref{sec:mean-field}. Outside of this interval it takes about a period of one small revolution for the system to go extinct.

\subsection{Short time dependence of the extinction probability}

In the short time limit, the semi-classical analysis presented in subsection \ref{ssec:semiclassics} should be accurate. Indeed, extinction in time $t\ll \tau_s$ is an exponentially rare event thats probability is convenient to represent as
$W= e^{-NS}$. Here  the action $S(G;t)$ is a solution of the Hamilton-Jacobi equation (\ref{HJ}) with the initial condition $S(0;0)=0$ and $G\to \infty$ at time $t$ . After the canonical transformation $(G,P_G)\to(X,P_X)$, the Hamiltonian acquires the form $H(X,P_X)=P_X^2$ and the classical equation of motion is $\dot X=2P_X$. We need its
solution reaching $X=X_0$ at time $t$. The corresponding action is
\begin{equation}\label{eqn:S}
S(X_0;t) = \frac{X_0^2}{4 t}\,,
\end{equation}
resulting in  the following form for the short time scale behavior of the extinction probability
\begin{equation}
\label{eqn:P-ext-short}
P_{ext} \propto e^{-N X_0^2/(4 t)}\,.
\end{equation}
This is exactly what was observed in stochastic simulations. The fit from Fig.~\ref{fig:hightcalc} gives $X_0=2.09$  for $\epsilon = 1$, while evaluating $X_0$ from Eq.~(\ref{eqn:Xdef}) results in $X_0 = 2.39$.
Again, the majority of the difference between these two values can be eliminated by using the value $X_0$ that is corrected for the finiteness of $N$, cf. Eq.~(\ref{eqn:Gextinction}). This was calculated in the previous section to be $X_0=2.08$, in much better agreement with the simulations. The $\epsilon$ dependence of the short time
scale $\tau_s=N X_0^2(\epsilon)/4$ follows from the dependence $X_0(\epsilon)$ discussed in the previous section.
Thus, one finds that away from the symmetric point $\epsilon=1$
\begin{equation}\label{eqn:taus-asymetric}
    \tau_s=2.2\, N \big(\mbox{max}\{\epsilon, 1/\epsilon\} \big)^{-2}\,.
\end{equation}
where the numerical constant is obtained through numerical integration of Eq.~(\ref{eqn:Xdef}).

One may argue that at the very smallest time scales the reduction of the initial Master equation (\ref{eqn:Master})
to the FP equation (\ref{eqn:FP}) may not be justified. As a result at such small times either Eq.~(\ref{eqn:P-ext-short}), or Eq.~(\ref{eqn:taus-asymetric}) may be violated.  Although this is a potentially valid concern, we were not able to go to sufficiently  short times (or sufficiently large $N$) to detect any sizable deviations of Monte Carlo results from analytical predictions, Eqs.~(\ref{eqn:P-ext-short}), (\ref{eqn:taus-asymetric}).

\section{\label{sec:discussion}Discussion}

We have investigated extinction due to intrinsic stochasticity  in the Lotka-Volterra model (\ref{eqn:reactions}).
To this end we have introduced two characteristic times: (i) the universal scale $\tau_l$, which characterizes exponential decay of survival probability at long times; (ii) the non-universal scale $\tau_s$ specific to the choice of initial condition close to the coexistence fixed point, which characterizes rise of the extinction probability
at short times. Since both these scales depend on the system parameters in exactly the same way and differ from each other only by a factor close to two, $\tau_s\approx 2.2\tau_l$, we shall restrict ourselves to discussions of the time $\tau_l$, which is independent on the choice of the initial conditions. All our results are valid in the asymptotic limit of large system size $N_{1,2}\gg 1$. 

\begin{figure}
\includegraphics[width=3in]{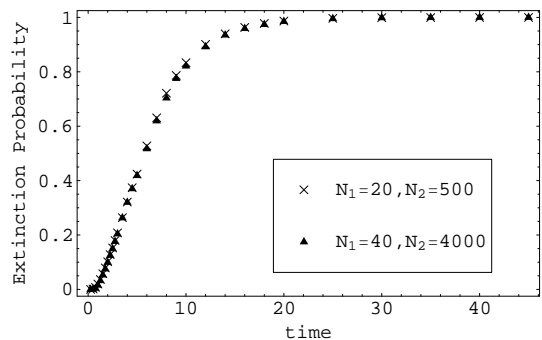}
\caption{\label{fig:two-models} Extinction probability of the two models: crosses $N=100$, $\epsilon=5$; triangles
$N=400$, $\epsilon=10$. In both cases $\tau_l\approx 4.0$, while $\tau_s\approx 8.8$.  }
\end{figure}

We consider first the asymmetric case. Recalling the definition of the parameters, Eq.~(\ref{eqn:N_eps}), and employing Eq.~(\ref{eqn:taul-asymetric}) one finds
\begin{equation}\label{eqn:taul-asymetric-1}
    \tau_l = \frac{N_s^{3/2}}{N_d^{1/2}}\,,
\end{equation}
where the size of the dominant population is $N_d=\mbox{max}\{N_1,N_2\}$, the size of the subdominant one is
$N_s=\mbox{min}\{N_1,N_2\}$, and time is measured in the natural units, which is the inverse frequency of the small cycles $1/\sqrt{\sigma\mu}$. This is a remarkable scaling relation, which predicts e.g. that the extinction time is shortened with increasing the size of the dominant population. Counterintuitively,  increasing abundance of dominant ``rabbits'' accelerates the extinction of subdominant ``foxes''! To check this prediction we performed stochastic modeling  of two prey-dominated models, which according to the scaling of Eq.~(\ref{eqn:taul-asymetric-1}) ought to go extinct
in the same relative time. The results are presented in Fig.~\ref{fig:two-models}. Since our method involved assumption that the angular motion is faster than the radial one, Eq.~(\ref{eqn:taul-asymetric-1}) may be trusted as long as
$\tau_l\gtrsim 1$, i.e. $N_d>N_s>N_d^{1/3}$. Outside of this interval of the parameters the extinction time is about one (in relative units).

Returning to the absolute scale of time and recalling that $N_1=\mu/\lambda$, while $N_2=\sigma/\lambda$, one may rewrite  the extinction time, Eq.~(\ref{eqn:taul-asymetric-1}), as
\begin{equation}\label{eqn:taul-asymmetric-2}
    \tau_l=\left\{ \begin{array}{ll}
    \mu/(\sigma\lambda)\,; &\quad \sigma>\mu\,,\\
    \sigma/(\mu\lambda)\,; &\quad \mu>\sigma\,.
    \end{array}\right.\,.
\end{equation}
The first line here is the prey-dominated case, while the second line the predator-dominated one. Again somewhat counterintuitively, increasing ``rabbits'' birth-rate accelerates their extinction in the ``fox''-dominated word.

In the symmetric case $N_d=N_s=N\gg 1$, we find
\begin{equation}\label{eqn:taul-symetric-1}
    \tau_l = 0.51\, N\to \frac{0.51}{\lambda} \,,
\end{equation}
where the first result is in the relative time scale, while the second in the absolute one. The linear scaling of the nearly symmetric  model with the system size is in agreement with the results of Ref.~\cite{Reichenbach}, obtained for a closely related cyclic model.  The factor close to
a half in comparison with the asymmetric case, Eq.~(\ref{eqn:taul-asymetric-1}), admits  a simple interpretation.
In the asymmetric case the diffusive current toward the extinction of the dominant population is exponentially smaller than that toward the extinction of subdominant species and may be neglected. In the symmetric case the two currents are exactly the same, making the extinction time twice shorter. How close to the symmetric point the system has to be
for Eq.~(\ref{eqn:taul-symetric-1}) to hold? Using Eq.~(\ref{eqn:iDG}) and taking characteristic value of $G$ from Eq.~(\ref{eqn:Gextinction}) one may estimate the corresponding interval of parameters as $|\epsilon-1|\lesssim 1/\ln N$, i.e. $N_d-N_s\lesssim N_d/\ln N_d$. This means that in the limit of large populations the symmetry condition is rather restrictive and a generic system most likely obeys the asymmetric scaling.

The natural extension of our study  is inclusion of spatial degrees of freedom. The spatial extension of the system is capable of stabilizing the system and increasing the extinction time~\cite{Durrett}. Even in a 2-site system, extinction time can be substantially longer than in the zero-dimensional case presented here~\cite{Abta}.
Understanding of such a stabilization mechanism is crucial for an accurate description of the phase transition
between the absorbing extinct phase and active coexistence phase, exhibited by the model on a thermodynamically large   d-dimensional lattice~\cite{Mobilia}.

We are indebted to M.~Dykman and B.~Meerson  for numerous illuminating discussions.  This research is
supported  by NSF Grants DMR-0405212 and DMR-0804266.

\appendix*

\section{\label{app:TD}Evaluation of $D(G)$ and $T(G)$}

In the  limit $G\ll \min\{\epsilon,1/\epsilon\}$, an orbit of constant $G$ is an ellipse. Both parameters, $D(G)$ and $T(G)$, may be found  exactly in this case
\begin{equation}
D(G)=2\pi G(\epsilon+1/\epsilon)\,;\quad\quad T(G)=2\pi\,.
\end{equation}
Equation~(\ref{eqn:1DFP}) takes the form
\begin{equation}
\partial_t{W}=\frac{\epsilon+1/\epsilon}{N}\frac{\partial}{\partial G}\left[G\,\frac{\partial W}{\partial G}\right].
\end{equation}
Changing variables as $G =  R^2$ near the mean-field fixed point gives the radial part of the two-dimensional diffusion equation with diffusion constant of $(\epsilon+1/\epsilon)/4N$
\begin{equation}
\partial_t{W} = \frac{\epsilon+1/\epsilon}{4N}\, \frac{1}{R}\, \frac{\partial}{\partial R}\left(R\, \frac{\partial W}{\partial R}\right)\,.
\end{equation}

The large $G$ limit can also be estimated.  The diffusive current $\vec{J}^D\cdot\hat{n}$ has two maxima corresponding to the minima  in one of the two species populations. These maxima are located at $Q_2=0$, $Q_1\approx -G-1/\epsilon$ and $Q_1=0$, $Q_2\approx -G-\epsilon$. Expanding near these two points the currents (\ref{eqn:JD1}) and (\ref{eqn:JD2}) and evaluating the integral  in Eq.~(\ref{eqn:DGdef}) one finds
\begin{eqnarray}
D(G)&=&\sqrt{\frac{\pi}{2}}\left({e+(1+\frac{1}{\epsilon^2})^{1/2+\epsilon^2}}\right)e^{\epsilon G} \nonumber\\
&+&\sqrt{\frac{\pi}{2}}\left({e+(1+\epsilon^2)^{1/2+1/\epsilon^2}}\right)e^{G/\epsilon}\,.
\end{eqnarray}
The majority of the orbital period is spent in the third quadrant.  In this quadrant, $\dot{Q_1} \approx -1$ and $Q_1$ varies from $\approx -G$ to $0$.  This gives for the orbital period
\begin{equation}
T(G)=G\,.
\end{equation}

\begin{figure}
\includegraphics[width=3in]{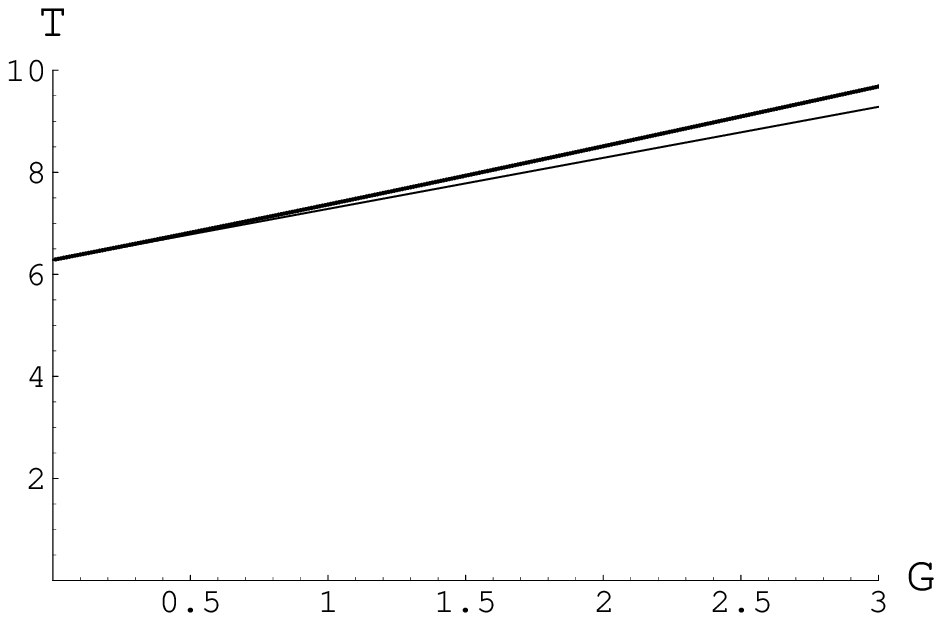}
\caption{\label{fig:tg} Numerically calculated $T(G)$ for $\epsilon = 1$ fit with analytically predicted $T(G)$}
\end{figure}

\begin{figure}
\includegraphics[width=3in]{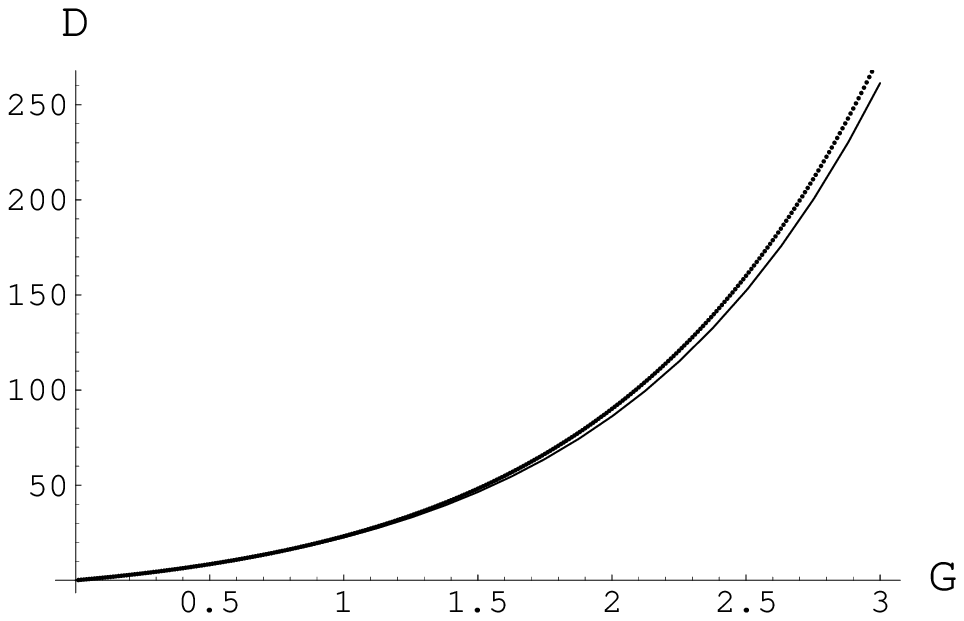}
\caption{\label{fig:dg} Numerically calculated $D(G)$ for $\epsilon = 1$ fit with analytically predicted $D(G)$}
\end{figure}
For the purposes of numerical diagonalization of the $\hat H$ operator we use the following
interpolating function  accurate in both the large and small $G$ limits
\begin{equation}\label{eqn:iTG}
T(G)=2 \pi + G\,.
\end{equation}
\begin{eqnarray}\label{eqn:iDG}
D(G) &=& 2\pi G(\epsilon + 1/\epsilon) \\
 &+& \sqrt{\frac{\pi}{2}}\left({e+(1+\frac{1}{\epsilon^2})^{1/2+\epsilon^2}}\right)(e^{\epsilon G}-\epsilon G - 1) \nonumber\\
&+&\sqrt{\frac{\pi}{2}}\left({e+(1+\epsilon^2)^{1/2+1/\epsilon^2}}\right)(e^{G/\epsilon}-G/\epsilon -1)\,. \nonumber
\end{eqnarray}
Figure \ref{fig:dg} shows the numerically calculated values for $D(G)$ fit with this interpolated $D(G)$. Figure \ref{fig:tg} shows the same for $T(G)$.

\bibliography{LV2}

\end{document}